\documentclass[twocolumn,preprintnumbers,prb]{revtex4}

\usepackage{graphicx}
\usepackage{dcolumn}
\usepackage{bm}

\begin{document}

\title{Design and Properties of a scanning EMR probe Microscope}

\author{S.A. Solin}
\email{solin@wustl.edu}
\affiliation{Center for Materials Innovation
And Department of Physics, Washington University in St. Louis
 1 Brookings Drive St. Louis, Missouri, 63130, USA}
\begin{abstract}
The design, fabrication, and predicted performance of a new type of magnetic 
scanning probe microscope based on the newly discovered phenomenon of 
extraordinary magnetoresistance (EMR) is described. It is shown that the new 
probe should advance the state of the art of both sensitivity and spatial 
resolution by an order of magnitude or more.
\end{abstract}
\maketitle
\section{Introduction}
During the past one and one half decades, various types of scanning probes 
have been developed to both image and measure the spatial variation of the 
magnetic fields near the surfaces of material systems such as high-density 
recording media,\cite{Companieh:2001} low and high temperature 
superconductors\cite{Bending:1999} and various types of other 
magnetic\cite{Howells:1999} and non-magnetic\cite{Nonmagnetic:1} systems. 
Understanding the micro structural details of the magnetic field 
distributions produced by such systems is crucial to the elucidation of the 
basic physical phenomena that govern their behavior. This understanding is 
greatly facilitated by high-resolution imaging and high sensitivity 
measurement of the magnetic field distribution associated with each system. 
The development of magnetic force microscopy (MFM)\cite{Volodin:1998} has 
greatly contributed to this endeavor but MFM has a few notable drawbacks. It 
measures the field gradient as opposed to the field itself. This complicates 
analysis and reduces the accuracy of the field distribution determination. 
In addition, the self-field of the MFM tip can be quite large, $\ge $ 1000 
Gauss, giving rise to an invasive probe in which the magnetic properties of 
the system under investigation are perturbed by the investigative tool.

To overcome the deficiencies in MFM Oral and co-workers developed a scanning 
Hall probe microscope (SHPM)\cite{Oral:1996} based on the GaAs/AlGaAs 
quantum well heterostructure and employed it to study vortecies in 
superconductors\cite{Crisan:2003} and the field distribution on or near the 
surface of insulating ferromagnets.\cite{Bending:1999} The SHPM is most 
attractive for low temperature measurements since its field sensitivity is 
proportional to the square root of the carrier mobility. That mobility can 
be of order $10^{6}$ cm$^{2}$/Vs for GaAs quantum wells at liquid He temperature but 
drops by a factor of 1000 or more at room temperature.\cite{Umansky:1997} 
Moreover, the three-dimensional spatial resolution of the SHPM is currently 
at best about 200 nm $\times$ 200 nm $\times$ 25 nm the latter being in the vertical 
direction. While this spatial resolution is adequate to distinguish key 
magnetic features (e.g. vortecies) in superconductors of interest and some 
features of ferromagnetic domains in insulators or metals, it is not 
sufficient to assess the details of the field distribution of such features 
without imposing deconvolution techniques that limit the accuracy of such 
determinations. However, the SHPM has a very short response time and it is 
constructed from non-magnetic material that is beneficial in minimizing the 
self-field of the probe.

The discovery of Extraordinary Magnetoresistance (EMR) by Solin and 
coworkers\cite{Solin:2000} and the fabrication of nanoscopic EMR field 
sensors\cite{Solin:2002} now provide the opportunity to advance the state 
of the art of semiconductor based scanning magnetic field probes by offering 
at least an order of magnitude higher sensitivity and an order of magnitude 
higher spatial resolution over a temperature range from liquid He 
temperatures \textbf{to room temperature} without sacrificing any of the 
intrinsic advantages of the SHPM. The route to this state of the art advance 
is effectively the replacement of the Hall probe in the SHPM with an 
appropriately designed and developed EMR probe thereby yielding a scanning 
EMR probe microscope or SEMRPM. Examples of measurements that will be made 
possible by the SEMRPM include but are not limited to:
\begin{itemize}
\item Imaging the bit field in ultra high density (TB/in$^{2})$ magnetic recording 
media
\item Ultrahigh resolution studies of current flow in quantum wires
\item Fault detection in nanocircuits
\item Probing new static and dynamic details of the vortex melting transition and 
phase diagram of high temperature superconductors
\end{itemize}
Here we describe the critical design criteria for an SEMRPM and evaluate the 
optimized performance properties which it can be expected to exhibit.

\section{Background -- EMR Physics, Macro- and Nano-structures}

There are two principal contributions to the magnetoresistance (MR) of any 
resistive device, namely a physical contribution and a geometric 
contribution.\cite{Popovic:1991} The physical contribution derives from the 
dependence of intrinsic material properties such as carrier concentration 
and carrier mobility on the applied magnetic field. The geometric 
contribution is an extrinsic property that depends on the shape of the 
device, the placement and geometry of the (metallic) contacts and, the 
placement and geometry of any inhomogeneities that may be present. The 
geometric contribution to the MR also depends on the intrinsic physical 
properties of the inhomogeneities relative to those of the host material, 
e.g. on the conductivity ratio.\cite{Tineke:1998} For most materials of 
current interest as MR sensors such as layered magnetic metals which exhibit 
giant MR (GMR)\cite{Egelhoff:1995} or tunneling MR (TMR)\cite{Mitra:2001} 
and the magnetic layered oxide manganites which exhibit colossal MR 
(CMR),\cite{Jin:1994} the physical contribution to the MR is dominant. 
However, Solin and his colleagues have recently shown that judiciously 
designed hybrid structures composed of a non-magnetic narrow-gap 
semiconductor matrix with high carrier mobility and a non-magnetic metallic 
inhomogeneity or shunt can exhibit a room temperature MR that is not only 
dominated by the geometric contribution but also attains room temperature 
values of order 1,000,000{\%} which is several orders of magnitude higher 
than what has been achieved with conventional GMR, TMR or CMR 
devices.\cite{Solin:2000} The new phenomenon was subsequently dubbed 
extraordinary MR or EMR.\cite{Solin:2003} The proof of principal 
demonstration of EMR was accomplished with symmetric 4-probe macroscopic van 
der Pauw (vdP) disc structures formed from Te-doped InSb (electron 
concentration $n = 2\times10^{17}$ cm$^{-3}$ and mobility $\mu=4.5\times10^{4}$ cm$^{2}$/Vs) containing a concentric
cylindrical metallic inhomogeneity as depicted in 
the inset of Fig. 1. 
\begin{figure}
\includegraphics[width=0.5\textwidth]{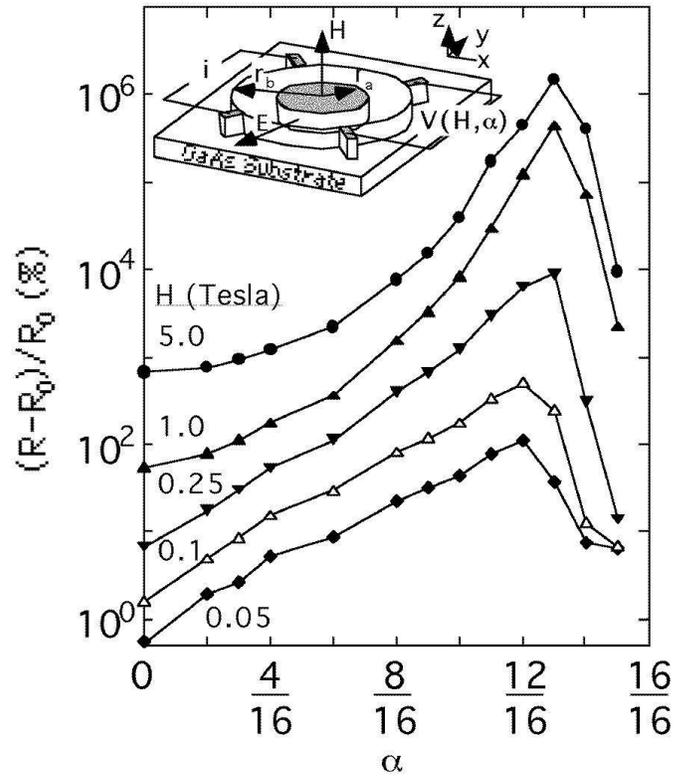}
\caption{The field-dependence of the magnetoresistance, 
$(R-R_{0})/R_{0}$, of a Te-doped InSb van der Pauw disk of radius $r_{b}$ in which 
is embedded a concentric right circular cylinder of Au of radius $r_{a}$. The 
filling factor is $\alpha =r_{a}/r_{b}$. Inset -- a schematic diagram of a composite 
disk with the current and voltage leads configured for the magnetoresistance 
measurement configuration.}
\end{figure}
Solin \textit{et al.} also showed that in general, 
$EMR(\Delta H,H_{bias})=[R^{eff}(\Delta H + H_{bias})-R^{eff}(H_{bias})]/R^{eff}(H_{bias})$
 where $H$ is the applied field normal to the plane of the device, 
$R_{eff}(H)$
 is the effective field-dependent resistance measured in a 4-probe 
configuration, 
$H_{bias}$
 is the bias field and 
$\Delta H$
 is the applied or signal field (not the field gradient). For small signals
\begin{eqnarray}
EMR(\Delta H\rightarrow 0,H_{bias})=\nonumber\\
\left[\frac{1}{R^{eff}(H_{bias})}\right]
\left[\frac{dR^{eff}(H)}{dH}\right]_{H_{bias}}\Delta H
\end{eqnarray}
where $[dR^{eff}(H)/dH]_{B_{bias}}$
is the current sensitivity. In the zero bias large signal but still low 
field limit, $\mu \Delta H \ll 1$,
\begin{eqnarray}
EMR(\Delta H, 0)=\frac{R^{eff}(\Delta H)-R_{0}^{eff}}{R_{0}^{eff}}=\nonumber\\
G_{S}(\Delta H)[\mu \Delta H]^{2}\pm G_{AS}(\Delta H)[\mu \Delta H].
\end{eqnarray}

Here 
$G_{S}(\Delta H)$ and $G_{AS}(\Delta H)$
are, respectively, symmetric and antisymmetric geometric factors which 
depend on the shape, location and physical properties of the conducting 
inhomogeneity and contacts while $R^{eff}(0)=R_0^{eff}$. [For the symmetric 
structure shown in the inset of Fig. 1. 
$G_{AS}(\Delta H)=0$.] Clearly, narrow-gap high mobility semiconductors such as InSb and InAs 
are choice materials for EMR devices.

The magnetotransport properties of the macroscopic vdP structure shown in 
the inset of Fig. 1. can be quantitatively accounted for using the above 
equations together with both finite element analysis\cite{Moussa:2001} and 
analytic techniques.\cite{Zhou:2001} However, the EMR phenomenon can be 
readily understood using a simple though non-intuitive classical physics 
analysis. The components of the magnetoconductivity tensor 
$\underline{\sigma ( H )}$ for the semiconductor are $\sigma 
_{xx} (\beta)=\sigma _{yy} (\beta)=\sigma /\left[ 
{1+\beta ^2} \right]$, $\sigma _{zz} (\beta)=\sigma $, and 
$\sigma _{xy} (\beta)=-\sigma \beta /\left[ {1+\beta ^2} 
\right]=-\sigma _{yx} (\beta)$ with $\beta =\mu H$ and all 
others being zero. If the electric field on the vertical surface of the 
inhomogeneity is $\vec {E}=E_x \hat {x}+E_y \hat {y}$, the current density 
is $\vec{J}=\underline{\sigma (H)}\vec{E}$. The electric field is everywhere normal to the 
equipotential surface of a highly conducting inhomogeneity. At $H = 0$, \underline{$\sigma (H)$}
 is diagonal so $\vec{J}=\sigma\vec{E}$ and the current flows into 
the inhomogeneity which acts as a \textit{short circuit}. At high $H$ ($\beta >1$), the 
off-diagonal components of \underline{$\sigma (H)$} dominate so
 $\vec{J}=(\sigma / \beta)[E_{y}\hat{x}-E_{x}\hat{y})]$
 and $\vec{J}\bot \vec{E}$.
Equivalently, the Hall angle between 
the electric field and the current density approaches 90$^{0}$, and the 
current becomes tangent to, i.e. deflected around, the inhomogeneity. Thus, 
the inhomogeneity acts as an \textit{open circuit}. The transition of the inhomogeneity from 
short circuit at low $H$ to open circuit at high $H$ results in a geometric 
enhancement of the MR of the semiconductor even if its resistivity 
(conductivity) is field-independent (i.e. the physical MR is zero). The MR 
increases with filling factor,
 $\alpha$, (see caption, Fig. 1.) 
because
 $R_{0}^{\alpha}$ decreases. However, when
  $\alpha$ 
becomes sufficiently large so that the low-field current flows mostly 
through the inhomogeneity, the MR will be that of the inhomogeneity itself, 
which for Au is negligibly small. Then an appreciable MR is only observed 
when $H$ is sufficient to deflect the current from the inhomogeneity such that 
the conductance through the metallic inhomogeneity is smaller than the 
conductance through the semiconductor annulus of thickness
 $r_{b}-r_{a}$. Clearly, the EMR effect results from orbital rather the spin 
degrees-of-freedom of the charge carriers.

Using conformal mapping methods Solin and coworkers showed experimentally 
and theoretically that macroscopic \textit{externally} shunted vdP plates were 
galvanomagnetically equivalent to the internally shunted disc shown in the 
inset of Fig. 1.\cite{Zhou:2001} They then faced the formidable 
challenge of scaling such EMR devices to the nanoscopic sizes required for 
ultra-high spatial resolution and high sensitivity detection of magnetic 
fields. To meet this challenge, they used an InSb/In$_{x}$Al$_{1-x}$Sb 
quantum well structure and state of the art suspended mask e-beam 
lithography incorporating a new type of resist, calixarine, to fabricate the 
structure shown in Fig. 2.\cite{Solin:2003} 
\begin{figure}
\includegraphics[width=0.5\textwidth]{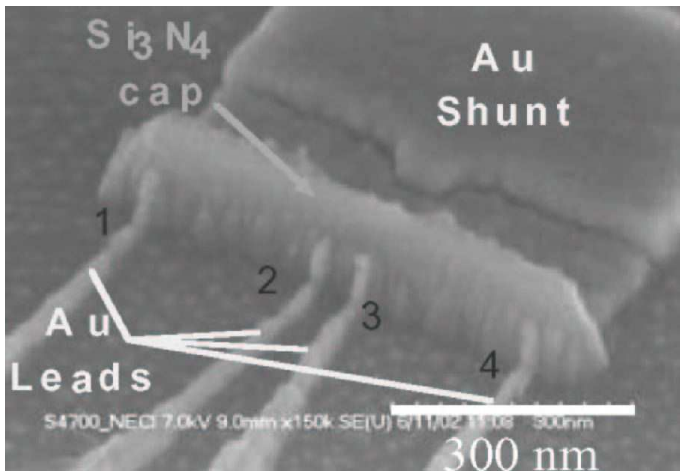}
\caption{An electron micrograph of a mesoscopic van der Pauw plate 
structure formed from an InSb/InAl$_{1-x}$Sb$_{x}$ quantum well. The current 
leads, voltage leads and external shunt are labeled as indicated.}
\end{figure}
Details of the fabrication 
method are provided elsewhere\cite{Pashkin:2000} with one exception. The 
leads and shunt on the device shown in Fig. 2 were insulated from the floor 
of the mesa containing the quantum-well by an Al$_{2}$O$_{3}$ layer that 
extended to within 50 nm of the mesa sidewall.

The field dependence of the room temperature magnetoresistance of the 
externally shunted nanoscopic EMR device shown in Fig. 2 is shown in Fig. 3. 
\begin{figure}
\includegraphics[width=0.5\textwidth]{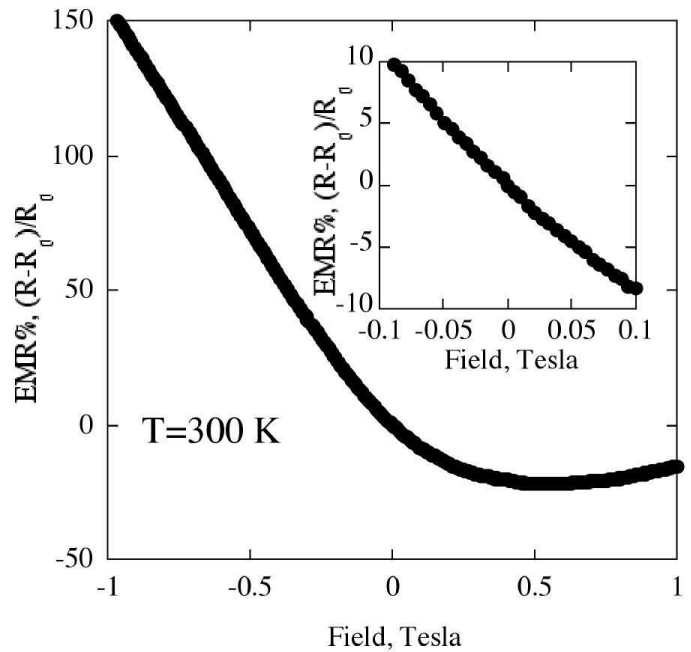}
\caption{The field dependence of the magnetoresistance of the mesoscopic 
van der Pauw plate structure shown in Fig.2. Bias current density, 
$5\times10^{3}$A/cm$^{2}$.}
\end{figure}
As can be seen, the EMR reaches values as high as 5{\%} at zero bias and a 
signal field of 0.05 T. To our knowledge, this is the highest room 
temperature MR level obtained to date for a patterned magnetic sensor with 
this spatial resolution. Moreover, with a modest bias field of 0.2 T 
corresponding to the zero-field offset\cite{Solin:1996} in Fig. 3, the 
measured EMR is 35{\%} at a signal field of 0.05 T. [The offset is 
associated with the asymmetric placement of the leads.] Also note that the 
device can be biased into a field region where the EMR response is linear 
with field, a feature that can simplify signal amplification. Equally 
significant is the fact that the current sensitivity, at a magnetic field 
bias of 0.2 T has a large measured value of 585 $\Omega$/T at room temperature. It 
is this figure that enters directly into the calculation of the signal to 
noise ratio as will be discussed below.

\section{Key Factors in the Design of an SEMRPM}
\subsection{Probe Materials}
Two materials and material systems will be of primary interest for 
nanoscopic EMR probes, namely InSb and InAs as well as quantum well 
structures based on those materials. InSb has already been shown to be an 
effective material for nanoscopic EMR sensors and thus for room temperature 
SEMRPM applications. But the Schottky barrier associated with the surface 
depletion layer in InSb\cite{Sze:1968} will limit its use in low 
temperature probes because of unacceptable increases in the shunt-mesa 
sidewall interface resistance. Barrier effects also preclude the use of 
GaAs/AlGaAs two dimensional electron gas (2Deg) structures, in nanoscopic 
EMR devices with mesa widths of order 30 nm, notwithstanding their huge 
mobility at low temperature. The material of choice for low temperature EMR 
probes will be InAs. It has already been shown to be effective in 
microscopic low temperature EMR devices.\cite{Ref:1} Moreover, InAS has an 
n-type surface accumulation layer with a high 2D carrier concentration of 
$1\times10^{12}$ cm$^{-2}$ and reasonable mobility of $2 \times 10^{4}$ cm$^{2}$/Vs 
both of which are relatively temperature independent below 
77 K.\cite{Wang:1992} Therefore, contact resistance to InAs with a number of 
metals is low even at low temperature.
\begin{figure}
\includegraphics[width=0.5\textwidth]{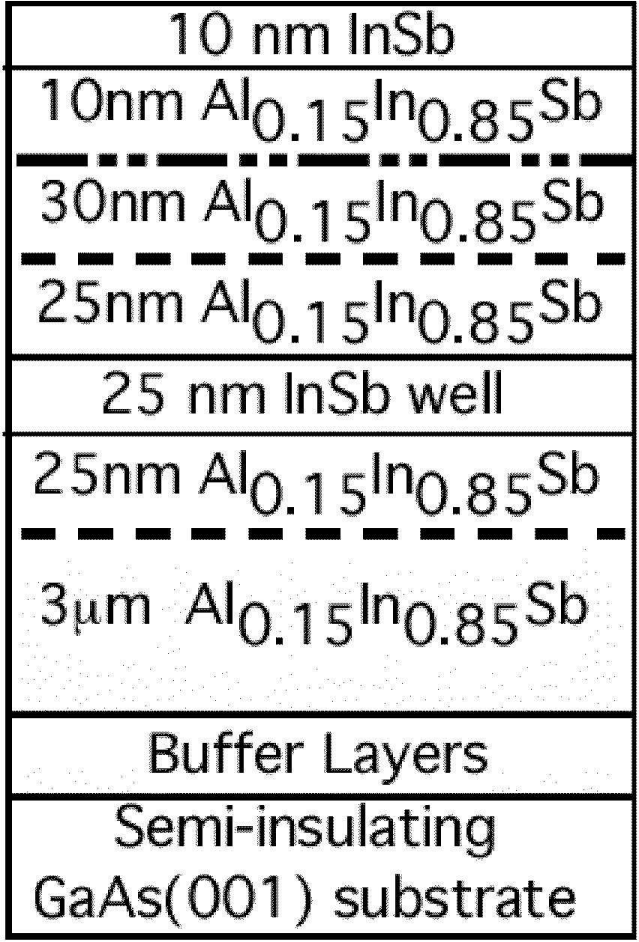}
\caption{ InSb/AlInSb quantum well structure with high mobility carriers in 
the InSb well at room temperature. The double (single) dashed line 
represents a $\delta $-doped Si layer with 2D concentration $9.5\times10^{11}$ 
cm$^{-2}$ $(2.75\times10^{11}$ cm$^{-2})$.}
\end{figure}

For a number of applications including the SEMRPM, it is desirable for the 
EMR sensor to be very thin in the vertical direction so that it has very 
high spatial resolution ($\sim $ 20 nm) in that direction and so that its 
active region can be positioned very close to the surface of the medium 
creating the field to be sensed. Unfortunately, although bulk thin film InSb 
on GaAs substrates can be prepared with room temperature electron mobility 
of order 50,000 cm$^{2}$/Vs, as one reduces the film thickness to values 
below about 1 $\mu $m, the mobility of currently available InSb films drops 
precipitously reaching a value of only 100 cm$^{2}$/Vs. at a film thickness 
of 0.1 $\mu $m.\cite{Parker:1989} Therefore, in order to provide the high 
carrier mobility that is required for high sensitivity at low fields in an 
EMR device [see Eq. (2)] the InSb mesa shown in Fig. 2 was etched from the 
InSb/In$_{x}$Al$_{1-x}$Sb quantum well structure shown schematically in Fig. 
4. This structure contains a 20 nm thick quantum well located about 90 nm 
from the top of the Si$_{3}$N$_{4}$ insulating cap layer [added to prevent 
shorting between the leads and the shunt.] The 2D concentration and the 
mobility of carriers in the well were measured to be 
$\tilde{n}=2.7\times10^{11}$ cm$^{-2}$
 and 
$\mu=2.3\times10^{4}$cm$^{2}$/Vs
 at room temperature. Note from Fig. 2 that the longitudinal resolution 
(along x) of the device is set by the spacing of the voltage probes because 
the shunt is designed to contact the opposite mesa sidewall along a length 
equal to the spacing between the voltage probes. Thus the volumetric 
resolution of the EMR device shown in Fig. 2 is 35 nm (the voltage probe 
spacing) $\times$ 30 nm (the width of the mesa) $\times$ 20 nm (the thickness of the 
quantum-well) along $x$, $y$ and $z$, respectively.

\subsection{Device Geometry and its Impact on Transport and Contact 
Resistance}

The room temperature mean free path of the carriers in an InSb quantum-well 
is 
$\ell=\hbar \sqrt{2 \pi \tilde{n}}(\mu / e)=200$ nm. Thus one would expect the transport in a nanostructure to be
ballistic in 
which case it can be shown that the expected EMR would be at least a factor 
of 5 lower than what is observed in Fig. 3. 
However, Solin \textit{et al.} have 
suggested that the transport is in-fact still diffusive as a result of the 
randomization of the carrier velocities due to scattering off of the rough 
mesa sidewalls\cite{Solin:2003} (see Fig. 2). The scattering process is 
enhanced because the roughness wavelength is of the order of the Fermi 
wavelength of the carriers 
$\lambda_{F}=\sqrt{2 \pi / \tilde{n}}= 48$ nm. Given the assumption of diffusive transport, the EMR of the nanoscopic 
device though noteworthy, is still about a factor of 20 lower than that 
obtained with the a macroscopic plate of the same geometry fabricated from 
thin film Te-doped InSb with a room temperature mobility of $4.5\times10^{4}$ 
cm$^{2}$/Vs.\cite{Zhou:2001} Part of this difference is due to the mobility 
difference thus yielding a reduction in EMR of a factor of 
$(4.5\times 10^{4}/2.3\times 10^{4})^{2} = 3.8$. The additional order of magnitude reduction derives from 
current leakage through the mesa floor (quantum well lower barrier) which 
carries a much higher proportion of the current than does the quantum well 
itself.

Rather than rendering the carriers diffusive by scattering off of the 
striated mesa sidewalls fluted current and voltage leads can be employed as 
depicted schematically in Fig. 5.

\begin{figure}
\includegraphics[width=0.5\textwidth]{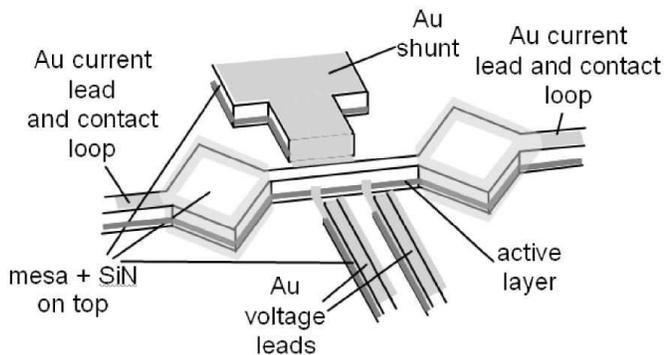}
\caption{Schematic diagram of EMR device with fluted current leads. 
[Features not to scale.]}
\end{figure}
The current leads will have dimensions 
much larger than the carrier mean free path. Thus current launched into the 
region of the shunt and voltage leads will already be diffusive owing to 
interactions in the fluted regions. Fluted leads will have the additional 
benefit of increasing the area of the current contacts and reducing contact 
resistance.

Of more concern is the contact resistance at the shunt interface. This 
diminishes the effectiveness of the shunt and reduces the EMR. For instance, 
the specific contact resistance of Au and similar metals with InSb at room 
temperature\cite{Sze:1968} is $\sim 10^{-7}$ $\Omega$ cm$^{2}$  while 
the contact area of the shunt used in the device shown in Fig. 2. is 
$9\times 10^{-12}$ cm$^{2}$. This 
yields a contact resistance of $1.1\times 10^{4}$ $\Omega$ 
that is about a factor of 10 larger than the intrinsic value of 
$R_{out}$
. Unfortunately, the shunt contact area cannot be increased without 
diminishing longitudinal resolution nor can the shunt be fluted without 
sacrificing transverse resolution. Moreover, the usual techniques of 
diffusing a metal such as In into the contact to eliminate the Schottky 
barrier or of adding a heavily doped interface layer between the quantum 
well side wall and the shunt are not applicable to nanoscopic structures of 
the type addressed here. If the SNR for the InSb -- Au system is inadequate 
for a particular probe application, the solution to the contact problem is 
to substitute InAs for InSb. This will be discussed below.

\subsection{Bias Electric Field and Maximum Input Current}

Three processes may, in principal, limit the maximum bias field one can 
apply to an EMR sensor. These are joule heating in the metal and/or 
semiconductor, electro- migration in the metal if it is an alloy structure 
and non-linear transport in the semiconductor. One can anticipate that 
nonlinear transport in the semiconductor will occur at bias fields well 
below those demarking the onset of the other processes (see below). The 
source of the nonlinear response in narrow-gap semiconductors is the 
scattering of carriers off of phonons and plasmons when those carriers are 
accelerated to sufficient energies to interact strongly with those 
entities.\cite{Rowe:2000} A direct consequence of the scattering by optic 
phonons is the reduction of the carrier mobility with increasing applied 
current or electric field. As a result the mobility of both InSb and InAs is 
essentially field independent (Ohmic) up to critical fields of order 
$E_{c}\sim 400$ V/cm
 above which, the mobility decreases precipitously with increasing field. 
The field dependence of the mobility of those materials can be well 
accounted for using a ``catchment'' model that incorporates the plasmon 
screening length as a key parameter and is applicable to diffusive 
transport.\cite{Rowe:2000}

From the information in Fig. 1, we can estimate the current density at which 
non-linear effects will occur in InSb. The critical current density that 
defines the onset of non-Ohmic response is given by 
$J_{c}=\sigma E_{c}=ne\mu E_{c}$
 where $\sigma$ is the material conductivity, $n$ is the carrier density (for electrons, the dominant carriers in InSb) and 
$\mu$ is the carrier mobility. If we assume typical carrier concentration of 
$n\sim 2\times 10^{17}$ cm$^{-3}$ 
and a typical mobility of $\mu \sim 3\times 10^{4}$ cm$^{2}$/Vs, then we find that 
$J_{c}\sim 3\times10^{5}$ Amps/cm$^{2}$
. Since critical currents for the onset of heat damage in InSb and InAs are 
an order of magnitude higher,\cite{Rowe:2000} the onset of nonlinear 
transport is a primary concern in establishing the maximum signal to noise 
ratio one can expect to obtain with an EMR sensor constructed from those 
materials.
\begin{table*}[ht]
\begin{center}
\begin{tabular}{|p{36pt}|p{84pt}|p{81pt}|p{81pt}|p{81pt}|p{81pt}|}
\hline
T(K)& 
Hall Probe$^{b}$ \par GaAs/AlGaAs \par 1$\mu $m $\times$ 1$\mu $m $\times$ 25nm& 
EMR Probe$^{c}$ \par InSb/AlInSb \par 1$\mu $m $\times$ 1$\mu $m $\times$ 25nm& 
EMR Probe$^{c}$ \par InSb/AlInSb \par 35nm$\times$30nm$\times$25nm& 
EMR Probe$^{d}$ \par InAs \par 1$\mu $m$\times$1$\mu $m$\times$25nm& 
EMR Probe$^{d}$ \par InAs \par 35nm$\times$30nm$\times$25nm \\
\hline
300& 
$4 \times 10^{-6}$ \par [$4 \times 10^{-6}$] \par {\{}0.016{\}}& 
$2.1 \times 10^{-8}$ \par [$7.4\times 10^{-5}$] \par {\{}0.300{\}}& 
$4.1 \times 10^{-6}$ \par [$2.2\times 10^{-6}$] \par {\{}0.009{\}}& 
$1.2 \times 10^{-8}$ \par [$1.3\times 10^{-4}$] \par {\{}0. 512{\}}& 
$2.4 \times 10^{-6}$ \par [$3.8\times 10^{-6}$] \par {\{}0.015{\}} \\
\hline
77& 
$3 \times 10^{-8}$ \par [$6 \times 10^{-5}$] \par {\{}0.24{\}}& 
NA& 
NA& 
$1.3 \times 10^{-9 e}$ \par [$2.0 \times 10^{-4}$] \par {\{}0.8{\}}& 
$2.6 \times 10^{-7}$ \par [$5.9 \times 10^{-6}$] \par {\{}0.024{\}} \\
\hline
4& 
$1 \times 10^{-8 e}$ \par [$6 \times 10^{-5}$] \par {\{}0.24{\}}& 
NA& 
NA& 
$9.8 \times 10^{-12 e}$ \par [$5.3 \times 10^{-4}$] \par {\{}2.1{\}}& 
$1.9 \times 10^{-9 e}$ \par [$1.6 \times 10^{-5}$] \par {\{}0.064{\}} \\
\hline
\end{tabular}
\label{tab1}
\end{center}
\caption{A comparison of the NEF(THz$^{-1/2})$, maximum bias 
current (A) [shown in square brackets] and self-field$^{a}$ (mT) {\{}shown 
in curly brackets{\}} of Hall and EMR probes at three different 
temperatures. $^{a}$ At a distance 50nm from the center of the active region of the sensor.
 $^{b}$ Data taken directly from
or computed from ref. 2. Oral \textit{et al.} (ref. 4) report a Hall probe with 250 nm lateral spatial resolution and a
field sensitivity of $\sim 3 \times 10^{-7}$ THz$^{-1/2}$ at $\sim $77 K but details were not available to allow
inclusion of this in Table I.
 $^{c}$  $n_{2D} = 5\times10^{11}$ cm$^{-2}$, $\mu _{300}=2.3\times10^{4}$ cm$^{2}$/Vs,
$E_{max} = 400$ V/cm, $H_{bias} = 0$. Assume fluted voltage and current contacts with negligible contact resistance.
$^{d}$ The temperature dependence of the mobility was computed using the Caughey {\&} Thomas formula. Parameters for the
surface accumulation layer: $n_{2D} = 1\times10^{12 }$ cm$^{-2}$, $\mu _{300}=2\times10^{4}$ cm$^{2}$/Vs, 
$E_{max} = 400$ V/cm. Assume $dR^{eff}/dH$ scales with $\mu ^{2}$. 
$^{e}$ Intrinsic value. Preamp noise limit at 300K is $\sim 3\times10^{-8}$.}
\end{table*}
\subsection{SEMRPM Performance}

The performance of a magnetic sensor is, of course, measured by the signal 
to noise ratio at the operating conditions under which it will be employed. 
We will propose sensor designs that result in diffusive transport at all 
temperatures of interest. For such EMR devices two noise sources are 
relevant, 1/f noise and thermal or Johnson noise. In this case, if the 
effective resistance is quadratic with field, e.g. 
$R^{eff}=R^{eff}_{0}\left[1+G\mu ^{2}(H-H_{0})^{2}\right]$
 where 
$H_{0}$
 is the zero-field offset, the voltage signal to noise ratio can be written 
in the following form:\cite{Weissman:1988}
\begin{eqnarray}
SNR(f)=\frac{I_{in}\left|\left(\frac{dR^{eff}}{dH}\right)_{H_{bias}}\right|\Delta H}
{\left[\left(\frac{V}{L}\right)^{2}\gamma\mu e R_{out}\frac{\Delta f}{f}+4kTR_{out}\Delta f\right]^{\frac{1}{2}}}=\nonumber\\
\frac{\left|El[(2G_{S}\mu ^{2}H_{bias}\pm G_{AS}\mu)\Delta H]\right|}
{\left[\frac{E^{2}\gamma l}{nwt}\frac{\Delta f}{f}+\frac{4kTl\Delta f}{nwte\mu}\right]^{\frac{1}{2}}}
\end{eqnarray}
where 
$I_{in}$
 is the input current, $V$ is input voltage, 
$L(l)$
 is the spacing of the current (voltage) leads, 
$\gamma$
 is the dimensionless Hooge parameter, $e$ is the electron charge, 
$f$
 is the operating frequency, 
$\Delta f$
 is the detection bandwidth, 
$k$
 is Boltzman's constant, $T$ is temperature in Kelvin, 
$R_{out}$
 is the two terminal resistance between the voltage probes including the 
contact resistance at the interface between the voltage probes and the 
device, 
$E=V/L$
 is the bias electric field, 
$l$
 is the voltage probe spacing, $n$ is the carrier (electron) density, $wt$ is the 
crossectional area for bias current flow and the other variables in Eq. (3) 
have been previously defined. The first term in each of the denominator 
brackets is the $1/f$ noise while the second term is the thermal noise. By 
equating these two terms we can deduce the crossover frequency 
$f_{c}=E^{2}\left(\gamma e \mu / 4kT\right)$
. For 
$f\gg f_{c}$
 thermal noise dominates and the SNR is frequency independent while for 
$f\ll f_{c}$
 $1/f$ noise dominates and the SNR is independent of the bias field.

It is useful at this point to estimate the crossover frequency for a 
nanoscopic EMR sensor. The relevant parameters are 
$\gamma \sim 10^{-8}$, 
$\mu\sim2.3\times 10^{4}$ cm$^{2}$/Vs, 
$E\sim 4\times 10^{2}$ V/cm (see discussion below) in which case 
$f_{c}\sim 400$Hz at 300K and $f_{c}\sim 30$KHz
 at 4K. Clearly, it is desirable to operate the EMR sensor at sufficiently 
high frequency to be in the thermal noise limited region. Moreover, since 
optimizing the controllable parameters in Eq. (3) to achieve minimal thermal 
noise collaterally minimizes the 1/f noise we focus here on the former. Note 
that $l$,$w$ and $t$ are set by the required three-dimensional resolution, and 
$\Delta H$ is set by the available signal so these parameters are deemed 
uncontrollable.

In the thermal noise and $1/f$ noise regimes the SNR increases as 
$n\mu^{5/2}$
 and 
$n\mu^{2}$
, respectively. Therefore, it is advantageous to maximize these products. 
Since for 
$n(T)>n_{c}(T)$
 mobility decreases\cite{Seeger:1985} with increasing 
$n$
 where 
$n_{c}(T)$
 is a critical concentration the SEMRPM will be designed to operate at or 
near 
$n_{c}(T)$
. For room temperature probes, 
$n_{c}(300$ K$)\sim5\times 10^{17}$ cm$^{-3}$
 for both InSb and InAs.\cite{Madulung:1991}

The SEMRPM performance can best be evaluated by comparing it to comparable 
probes based on the Hall effect in GaAs/AlGaAs heterostructures since the 
two sensor types will share many similarities. The figure of merit 
appropriate to such a comparison is the noise equivalent field 
(NEF)\cite{Deeter:1993} that is obtained from Eq. (3) by setting 
$SNR(f)=1$
 and solving for the resultant value of
$\Delta H / \sqrt{\Delta f}$
. The thermal noise limit NEF figures of an EMR probe fabricated from an 
InSb quantum well structure and from InAs bulk thin films are compared to 
corresponding values of a Hall probe in Table I. 
Also shown in that table 
are the self-field of the probe (e.g. the field near the probe produced by 
its own bias current) at a distance of 50 nm from the center of its active 
region. To facilitate direct comparison, the performance of each probe is 
calculated for a standard set of dimensions 1$\mu $m $\times$ 1$\mu $m $\times$ 25 nm 
using reported/measured maximal current densities. For the Hall probe, key 
data is taken directly from ref. 2 and references therein. For the InSb 
probe, the current sensitivity of 585 $\Omega$/T is taken from the data of Fig. 
3. For InAs the current sensitivity is assumed to scale as $\mu ^{2}$ and the 
temperature dependence of $\mu $ is obtained from the Caughey - Thomas 
formula.\cite{Caughey:1967}

As can be seen from Table I, the EMR probe offers significant potential 
advantages over the Hall probe in sensitivity, spatial resolution and 
self-field strength. Moreover, like the Hall probe, the EMR probe should be 
capable of very wide band width operation leading to attractive signal to 
noise ratios. While the data of Table I appears to favor the InAs system 
over the InSb system for EMR probe applications at all temperatures, the 
latter appears to offer greater opportunity for developing high mobility 
materials for room temperature applications. Therefore, it is proposed that 
both systems be developed in parallel in order to insure the development of 
a superior EMR probe. 

\subsection{SEMRPM Fabrication and Construction}

Wafers will be fabricated into the mesoscopic structures appropriate to the 
SEMRPM application and instrumentation. The fabrication process which has 
been described in detail elsewhere,\cite{Solin:2002} will involve 
application of a Si$_{3}$N$_{4}$ overcoat to the material wafer, state of 
the art e-beam lithography using calixarene resist, reactive ion etching and 
suspended mask metallization. The EMR probe will be coupled to the 
piezoelectric scanner tube of a commercial low temperature scanning 
tunneling microscope housed in an superconducting solenoid capable of 1.5K 
-- 325K operation at fields up to 10T. The design is similar to one 
developed by Bending and coworkers\cite{Bending:1999} that is popular for 
SHPM's and allows for convenient simultaneous acquisition of both the 
(standard DC\cite{Bending:1999} or AC\cite{Marchevsky:2001}) EMR probe 
and the STM signals thus providing both topographic and magnetic images.

\section{Summary and Conclusions}
We have shown that the SEMRPM, once constructed according to the design 
criteria described above, should significantly advance the state of the art 
of magnetic nanoprobe technology. A program has been undertaken by the 
author to fabricate the SEMRPM and to employ it to address several problems 
in basic and applied science. Results derived from this program will be 
reported in the near future.

\section{Acknowledgments}
Useful discussions with A.C. H. Rowe, D.R. Hines and L. Cohen are gratefully 
acknowledged. This work was supported by the NSF under grant {\#} 
ECS-0329347.


\begin{thebibliography}{3}
\bibitem{Companieh:2001}. A. Companieh, R. Eaton, R. S. Indeck, and M. Moser, IEEE Trans. on Magn., (2001).
\bibitem{Bending:1999}. S.J. Bending, \textit{Advances in Physics} \textbf{48}, No. 4 (1999).
\bibitem{Howells:1999}. G.D. Howells, A. Oral, S.J. Bending, S.R. Andrews, P.T. Squire, P. Rice, A. de Lozanne, J.A.C. Bland, I. Kaya and M. Henini, \textit{J. Magnetism and Mag. Matls}. \textbf{917}, 196 (1999).
\bibitem{Nonmagnetic:1}. Nonmagnetic system such as quantum wire with current.
\bibitem{Volodin:1998}. A.~Volodin, K. Temst, C.~Van~Haesendonck and Y. Bruynseraede, \textit{Appl. Phys. Lett.} \textbf{73}, 1134 (1998).
\bibitem{Oral:1996}. A. Oral, S.J. Bendin and M. Henini, \textit{Appl. Phys. Lett.} \textbf{69}, 1324 (1996).
\bibitem{Crisan:2003}. A Crisan ,A Pross, R G Humphreys and S Bending, \textit{Supercond. Sci. Technol}. \textbf{16, }695 (2003).
\bibitem{Umansky:1997}. V.Umansky, R.de-Picciotto, and M.Heiblum, \textit{Appl.Phys.Lett.}\textbf{ 71}, 683 (1997)
\bibitem{Solin:2000}. S.A. Solin, Tineke Thio, D.R. Hines and J.J. Heremans, Science \textbf{289}, 1530 (2000).
\bibitem{Solin:2002}. S.A. Solin, D.R. Hines, J.S. Tsai, Yu. A. Pashkin, S.J. Chung, N. Goel and M.B. Santos, \textit{Appl. Phys. Letters} \textbf{80}, 4012 (2002).
\bibitem{Popovic:1991}. R.S. Popovic, "Hall effect device\textit{s"}, (Adam Hilger, Bristol, 1991).
\bibitem{Tineke:1998}. Tineke Thio and S.A. Solin, \textit{Appl. Phys. Lett.}, \textbf{72}, 3497 (1998).
\bibitem{Egelhoff:1995}. W.F. Egelhoff, Jr., \textit{et al}., \textit{J. Appl. Phys.} \textbf{78}, 273 (1995).
\bibitem{Mitra:2001}. C. Mitra, P. Raychaudhuri, G. Kobernik,, K. Dorr, K.H. Muller, L. Schultzand and R. Pinto, \textit{Appl. Phys. Lett.} \textbf{79}, 2408(2001)
\bibitem{Jin:1994}. S. Jin, M. McCormack, T.H. Tiefel and R. Ramesh, \textit{J. Appl. Phys.} \textbf{76}, 6929 (1994).
\bibitem{Solin:2003}. S.A. Solin, D.R. Hines, J.S. Tsai, Yu. A. Pashkin, S.J. Chung, N. Goel and M.B. Santos, \textit{Appl. Phys. Letters} \textbf{80}, 4012 (2002).
\bibitem{Moussa:2001}. J. Moussa, L. R. Ram-Mohan, J. Sullivan, T. Zhou,$^{ }$ D. R. Hines,$^{ }$ and S. A. Solin, \textit{Phys Rev.} \textbf{B64}, 184410 (2001).
\bibitem{Zhou:2001}. T. Zhou, D.R. Hines and S.A. Solin, \textit{Appl. Phys. Lett.} \textbf{78}, 667 (2001).
\bibitem{Pashkin:2000}. Yu.~A.~Pashkin, Y.~Nakamura, and J.~S.~Tsai, \textit{Appl. Phys. Lett.} \textbf{76}, 2256 (2000).
\bibitem{Solin:1996}. S.A. Solin, Tineke Thio, J.W. Bennett, and D.R. Hines M. Kawano, N. Oda, and M. Sano, \textit{Appl. Phys. Lett}. \textbf{69}, 4105 (1996).
\bibitem{Sze:1968}. S. Sze, "Physics of Semiconductor Devices", (Wiley, New York, 1968).
\bibitem{Ref:1}. Ref to grundler's work in InAs Microscopic structures.
\bibitem{Wang:1992}. P.D. Wang \textit{et al.}, \textit{Semiconductor Science {\&} Tech.} \textbf{7}, 767 (1992).
\bibitem{Parker:1989}. S.D. Parker, \textit{et al}., \textit{Semicond. Sci. Technology} \textbf{4}, (1989) 663.
\bibitem{Rowe:2000}. A.C.H. Rowe, C. Gatzke, R.A. Stradling and S.A. Solin, \textit{Appl. Phys. Lett.}\textbf{ 76}, 1902 (2000)
\bibitem{Weissman:1988}. M.B. Weissman, \textit{Rev. Mod. Phys.} \textbf{60}, 537 (1988).
\bibitem{Seeger:1985}. K. Seeger, "Semiconductor Physics", (Springer-Verlag, Berlin, 1985, 3rd edition)
\bibitem{Madulung:1991}. O. Madulung, Ed., "Semiconductors: Group IV Elements and III --V Compounds", (Springer-Verlag, New York, 1991).
\bibitem{Deeter:1993}. M.N. Deeter, G.W. Day; T.J. Beahn; M. Manheimer, \textit{Elec. Lett.} \textbf{29} 933 (1993).
\bibitem{Caughey:1967}. D.~Caughey and R.Thomas, \textit{Proc.IEEE}, \textbf{52}, 2192 (1967).
\bibitem{Marchevsky:2001}. M. Marchevsky, M. J. Higgins and S. Bhattacharya, Nature \textbf{409}, 591 (2001).\par 
\end{thebibliography}
\end{document}